\documentclass[twocolumn,trackchanges]{aastex701}
\usepackage{amsmath}
\usepackage{bm}
\usepackage[title]{appendix}
\usepackage{xcolor}

\begin{document}

\title{Enhancing the Interpretability of Radial Velocity Analysis Using Leave-one-out Cross-validation}

\correspondingauthor{Matthew C.\ Nixon}
\email{matthewnixon@asu.edu}

\author[0000-0001-8236-5553]{Matthew C.\ Nixon}
\altaffiliation{51 Pegasi b Fellow}
\affiliation{School of Earth and Space Exploration, Arizona State University, Tempe, AZ 85287, USA}
\email{matthewnixon@asu.edu}

\author[0000-0003-4508-2436]{Ritvik Basant}
\affiliation{Department of Astronomy \& Astrophysics, University of Chicago, Chicago, IL 60637, USA}
\email{}

\author[0000-0002-4671-2957]{Rafael Luque}
\affiliation{Department of Astronomy \& Astrophysics, University of Chicago, Chicago, IL 60637, USA}
\affiliation{Instituto de Astrofísica de Andalucía (IAA-CSIC), Glorieta de la Astronomía s/n, 18008 Granada, Spain}
\email{}

\author[0000-0003-0156-4564]{Luis Welbanks}
\affiliation{School of Earth and Space Exploration, Arizona State University, Tempe, AZ 85287, USA}
\email{}

\author[0009-0003-2535-5641]{Pierrot Lamontagne}
\affiliation{Instituto de Astrofísica de Andalucía (IAA-CSIC), Glorieta de la Astronomía s/n, 18008 Granada, Spain}
\email{}

\begin{abstract}

Radial velocity (RV) observations are reaching increasingly high precision, enabling the targeting of low-amplitude signals. Assessing results obtained at the limits of our capabilities requires understanding the relationship between data, models and inference. Leave-one-out cross-validation (LOO-CV) has become an increasingly popular tool for interpretable model criticism in astronomical data analysis. In this work, we demonstrate how LOO-CV can provide insight into the RV analysis of exoplanetary systems through three case studies. We first investigate the anomalously high mass of HD~119130~b inferred from CARMENES RVs, later revised significantly downward after follow-up observations with other instruments. We show that this high mass was strongly driven by two individual measurements, whose removal significantly reduces the tension between data sets. Next, we use LOO-CV as a model selection tool for the GJ~4276 system. Here, LOO-CV prefers an eccentric single-planet model over a two-planet model, and identifies the orbital phases at which follow-up observations would best discriminate between them. Finally, we extend LOO-CV to models with correlated noise, applying it to the GJ~357 system, whose RVs are modeled with a Gaussian process. We find no predictive support for a fourth planetary signal that is moderately favored by the Bayesian evidence, strengthening the interpretation of a three-planet architecture previously reported in the literature. We provide practical guidance for implementing LOO-CV in RV analysis, together with an online tutorial. Such methods will become ever more important as future RV surveys push the limits of planet detection and characterization.

\end{abstract}


\section{Introduction} \label{sec:intro}

Radial velocity (RV) techniques have achieved remarkable progress in recent years, enabling the detection and mass measurement of an increasingly diverse range of exoplanets \citep{Fischer2016,Crass2021,Burt2025}. One of the major goals of the field is to precisely and accurately measure the masses of terrestrial exoplanets, since this is a crucial step towards further atmospheric and interior characterization \citep{Batalha2019b}. Although the smaller signal size imprinted by these lower-mass planets presents a challenge, we are now capable of discovering and measuring the masses of terrestrial planets orbiting M dwarf stars \citep[e.g.,][]{Anglada-Escude2016, ProxCen2022, Barnard2025, Brady2025} and the field is pushing towards mass measurements of Earth-Sun analogs \citep[e.g.,][]{Figueira2025, Dumusque2026}.

As the field builds towards Extreme Precision RV architectures and capabilities, several challenges must be overcome to ensure further progress, such as understanding the intrinsic variability of host stars, developing community coordination networks and observing strategies among telescopes to overcome sparse data and sub-optimal sampling, and successfully leveraging knowledge gained from existing observations \citep{Crass2021}. Additionally, model interpretability is key when trying to assess low-amplitude signals, which are especially susceptible to biases in mass and eccentricity estimates \citep{Zakamska2011,Hara2019,Luhn2023}. Finally, methods used to detect planets from RV observations, such as Bayesian model comparison, may suffer from shortcomings such as a strong dependence on priors, especially in the low signal-to-noise regime \citep{HaraFord2023, Welbanks2026}. Any path forward must include the development of new methodologies for statistical assessment and validation, including optimization of follow-up studies and refinement of planetary (e.g., mass) and orbital (e.g., eccentricity) parameters.

One such methodology for statistical assessment is the family of techniques known as cross-validation. In general, cross-validation methods evaluate how well a model trained on some data will perform when applied to new, unseen data. A number of previous works have demonstrated the value of applying cross-validation techniques to RV data analysis. \citet{Cloutier2017} used time-series cross-validation, in which model parameters are optimized on a chronological subset of the observations and the likelihood is evaluated on the next measurement in the sequence, both to select between stellar activity models and to establish the number of planets in the K2-18 system. \citet{Cloutier2019} and \citet{Radica2022} instead omitted individual measurements in turn and recomputed the significance of a candidate planetary signal in a periodogram of the remaining data, thereby identifying measurements whose inclusion suppresses that signal. \citet{CollierCameron2021} applied cross-validation to determine the number of principal components of the observed spectral line-shape variations to retain when modeling stellar activity, and \citet{Hara2025} used it to compare models with different numbers of activity indicators. \citet{Jones2022} evaluated cross-validation scores on withheld blocks of consecutive observations, and \citet{Blunt2023} withheld the measurements of an entire instrument. 

Within the family of cross-validation techniques, leave-one-out cross-validation \citep[LOO-CV;][]{Vehtari2017} has become increasingly prevalent in numerous areas of astronomy \citep[e.g.,][]{Morris2021,Neil2022,McGill2023,Welbanks2023, Welbanks2024, Nixon2024, Bello2025, Zhang2025}. LOO-CV involves removing each individual data point one-at-a-time, and measuring how well each point is predicted by the model trained on the remaining data. This provides insight into how individual data points can drive parameter inference, as well as providing an overall metric for the out-of-sample predictive performance of a given model. In this paper, we explore the benefits of applying LOO-CV to RV data analysis. Our approach differs from previous work in three main respects:

\begin{itemize}
    \item The quantity we compute for each withheld measurement is its out-of-sample predictive density under each model, rather than the effect of its removal on an in-sample detection statistic \citep{Cloutier2019,Radica2022}, which allows the predictive power of competing planetary architectures to be compared at the per-data-point level as well as in aggregate.
    \item We employ methods to allow these quantities to be obtained without refitting the model once per measurement (see Section \ref{sec:methods}), resulting in a method that is highly computationally efficient.
    \item We extend the method to likelihoods that do not factorize over the data, which is required whenever the RV model includes a Gaussian process (GP), as is commonly necessary in order to account for stellar activity.
\end{itemize}

Overall, the goal of this paper is to address the following question: given a set of RV measurements and one or more candidate models, which individual measurements are driving the inferred planetary parameters and any preference between models, and are those inferences robust enough to survive the addition of further data? We present the underlying methodology in Section \ref{sec:methods}. In Section \ref{sec:results} we present three case studies demonstrating possible applications of LOO-CV to RV analysis: (1) diagnosing the accuracy of limited data sets, using HD~119130~b as a case study; (2) assessing model preference between high-eccentricity and multi-planet solutions, using the example of the GJ~4276 system; and (3) assessing the significance of a marginal planetary signal in the presence of correlated noise, using the example of the GJ~357 system. In case (2), we also show how LOO-CV can be used to optimize follow-up observations. We discuss practical advice for implementation and possible future directions in Section \ref{sec:discussion}.


\section{Methods} \label{sec:methods}

\subsection{Radial velocity analysis}

For all of our RV analyses, we use \texttt{juliet} \citep{Juliet}, a flexible wrapper of multiple libraries that allow for the analysis of transit and RV data sets in the presence of correlated noise using Nested sampling algorithms. In particular, for this work, we use \texttt{RadVel} \citep{2018PASP..130d4504F} for the RV modeling and \texttt{dynesty} \citep{Speagle2020} for sampling. We revisit previous analyses of the HD 119130 \citep{Luque2019b}, GJ 4276 \citep{Nagel2019} and GJ 357 \citep{Luque2019} systems, fitting for the relevant orbital parameters ($P$, $t_0$, $e$, $\omega$, $K$) for different numbers of planets as specified in each case study. Velocity offsets and white noise terms are added for each instrument using wide priors, as is standard practice in RV analyses. Details of the priors for each analysis are given in Tables \ref{Table:HD119130}---\ref{Table:GJ357}.

\subsection{Leave-one-out cross validation} \label{subsec:loocv}

The LOO-CV method has been described in detail in numerous previous papers \citep[e.g.,][]{Vehtari2017,Welbanks2023,Challener2023}. Here, we summarize the method and refer the reader to the references above for a complete account.

Suppose a model $\mathcal{M}$ has been fit to a data set consisting of data points $d_i$. For each point, we wish to calculate $\mathrm{elpd}_{\mathrm{LOO}, \mathcal{M}, i}$, the expected log pointwise predictive density:

\begin{align}
    \mathrm{elpd}_{\mathrm{LOO},\mathcal{M}, i} =  \log p \left( d_i | d_{-i}, \mathcal{M} \right),
\end{align}

\noindent where $p \left( d_i | d_{-i}, \mathcal{M} \right)$ is the posterior predictive density of data point $d_i$ under model $\mathcal{M}$ trained on the remainder of the data set after $d_i$ has been removed ($d_{-i}$). This is calculated by taking the expectation of the likelihood of $d_i$ with respect to the posterior of $\mathcal{M}$ trained on $d_{-i}$:

\begin{align}
    p \left( d_i | d_{-i}, \mathcal{M} \right) = \int p \left( d_i | \theta, \mathcal{M} \right) p \left( \theta | d_{-i}, \mathcal{M} \right) \mathrm{d} \theta,
    \label{eq:prob}
\end{align}

\noindent where $\theta$ represents the parameters of $\mathcal{M}$. In practice this expectation is evaluated from a set of $S$ posterior samples $\theta^s$ as 

\begin{align}
    \mathrm{elpd}_{\mathrm{LOO},\mathcal{M},i} \approx \log \left[ \frac{1}{S} \sum_{s=1}^{S} p(d_i|\theta^s,\mathcal{M}) \right].
    \label{eq:exact_loo}
\end{align}

\noindent The total expected log posterior predictive density of the model can then be calculated by taking the sum over all $N$ data points:

\begin{align}
    \mathrm{elpd}_{\mathrm{LOO},\mathcal{M}} &= \sum_{i=1}^{N} \mathrm{elpd}_{\mathrm{LOO},\mathcal{M}, i} \, .
\end{align}

These metrics can be used to compare the predictive performance of two competing models, $\mathcal{M}_1$ and $\mathcal{M}_2$, by calculating

\begin{align}
    \Delta\mathrm{elpd}_{\mathrm{LOO},\mathcal{M}_1,\mathcal{M}_2} &= \mathrm{elpd}_{\mathrm{LOO},\mathcal{M}_1} - \mathrm{elpd}_{\mathrm{LOO},\mathcal{M}_2} \, ,
\end{align}

\noindent and performance at the per-data-point level can be compared by calculating $\Delta\mathrm{elpd}_{\mathrm{LOO},\mathcal{M}_1,\mathcal{M}_2, i}$.

Computing $\mathrm{elpd}_{\mathrm{LOO},\mathcal{M}}$ as described above requires refitting the model $N$ times, which may prove computationally prohibitive for large data sets. In these cases, Pareto-smoothed importance sampling (PSIS) can be used to approximate the values of $\mathrm{elpd}_{\mathrm{LOO}, \mathcal{M}, i}$ \citep{Vehtari2015,Vehtari2017}. PSIS draws $S$ posterior samples from a model fit to the entire data set, and for each sample $s$ and data point $i$ calculates an importance ratio $r^s_i$:

\begin{align}
    r^s_i = \frac{1}{p \left( d_i|\theta^s, \mathcal{M} \right)} \propto \frac{p \left( \theta^s | d_{-i}, \mathcal{M} \right)}{p \left( \theta^s | d, \mathcal{M} \right)}. \label{eq:weights}
\end{align}

To avoid possible domination by extreme values, the raw importance ratios are fit with a Pareto distribution \citep{zhang2009} to create smoothed importance weights $w^s_i$, and the samples are then re-weighted to approximate the effect of leaving out a data point:

\begin{align}
    p \left( d_i | d_{-i}, \mathcal{M} \right) &= \int p \left(d_i | \theta, \mathcal{M} \right) \frac{p \left( \theta | d_{-i}, \mathcal{M} \right)}{p \left( \theta | d, \mathcal{M} \right)} p \left( \theta | d, \mathcal{M} \right) \mathrm{d} \theta \nonumber \\
    & \approx \frac{\sum_{s=1}^{S} w_i^s p \left( d_i | \theta^s, \mathcal{M} \right) }{\sum_{s=1}^{S} w_i^s }. \label{eq:psis}
\end{align}

The estimated shape parameter $k$ of the fitted Pareto distribution serves as a performance indicator of the PSIS method, with $k \geq 0.7$ indicating that the approximation is unreliable for a given data point\footnote{If $k>0.5$, this indicates that the variance of the raw importance weights is effectively infinite, meaning that the estimator is dominated by a few extreme weights. However, extensive empirical testing has shown that the PSIS approximation remains reliable for $0.5 < k < 0.7$ \citep{Vehtari2015}.}. For those data points where $k \geq 0.7$, a full refit to the data with that point removed is required. A more detailed explanation of PSIS can be found in \citet{Vehtari2015}.

Since LOO-CV estimates out-of-sample predictive performance using a finite data set, both $\mathrm{elpd}_{\mathrm{LOO},\mathcal{M}}$ and $\Delta\mathrm{elpd}_{\mathrm{LOO},\mathcal{M}_1,\mathcal{M}_2}$ carry a standard error (SE). This standard error reflects the sampling uncertainty of the data: because we have only a finite set of RV measurements, the total $\mathrm{elpd}_{\mathrm{LOO},\mathcal{M}}$ is itself an estimate, and a different draw of measurements from the same system would yield a slightly different value. Treating the $N$ pointwise values as a sample from the underlying data-generating distribution \citep{Vehtari2017} gives

\begin{align}
    \rm{SE} \left( \mathrm{elpd}_{\mathrm{LOO},\mathcal{M}} \right) &= \sqrt{N \, \rm{Var} \left( \mathrm{elpd}_{\mathrm{LOO},\mathcal{M}, \textit{i}} \right)} \,  , \label{eq:se_1} \\
    \rm{SE} \left( \Delta\mathrm{elpd}_{\mathrm{LOO},\mathcal{M}_1,\mathcal{M}_2} \right) &= \sqrt{N \, \rm{Var} \left( \Delta\mathrm{elpd}_{\mathrm{LOO},\mathcal{M}_1,\mathcal{M}_2, \textit{i}} \right)} \, , \label{eq:se_2}
\end{align}

\noindent where $\rm{Var} (\cdot)$ is the variance operator. Reporting $\Delta\mathrm{elpd}_{\mathrm{LOO},\mathcal{M}_1,\mathcal{M}_2}$ in units of its SE therefore indicates whether a difference in predictive performance between two models is larger than the uncertainty that can be attributed to the finite size of the data set. We note that Equations \ref{eq:se_1} and \ref{eq:se_2} treat the $N$ pointwise contributions as independent. Where the noise is correlated (Section \ref{subsec:corrnoise}) the pointwise terms are themselves correlated, and the resulting SE should be regarded as approximate.

\subsection{Correlated noise}
\label{subsec:corrnoise}

The methods described above can be implemented straightforwardly in cases where each data point can be assumed to have independent noise properties. This assumption is made in Equation \ref{eq:prob}, and it holds for the case studies in Sections \ref{subsec:hd119} and \ref{subsec:gj4276}, but not in general. Under the assumption of Gaussian noise, one simply needs to calculate the log likelihood function for each data point:

\begin{equation}
    \log p(d_i| \theta, \mathcal{M}) = -\frac{1}{2} \left[ \log 2 \pi \sigma_i^2 + \frac{(d_i-m_i)^2}{\sigma_i^2} \right],
\end{equation}
where $\sigma_i$ is the measured uncertainty on data point $i$ and $m_i$ is the $i^{\rm th}$ model point. However, effects such as stellar activity are known to impart correlated noise in many RV data sets \citep[e.g.,][]{Haywood2014,Rajpaul2015,Aigrain2023}. It is therefore common to include a GP component \citep[][]{Rasmussen2006} when fitting RV observations \citep[e.g.,][]{Juliet,Barragan2022,Luque2019,Barnard2025}.

In cases where a GP has been used to model RV data, LOO-CV can still be employed, instead using the exact conditional predictive probability of each left-out point, given a specific set of hyperparameters (${\theta}_{\rm GP}$) and the rest of the data set \citep{Burkner2021}:

\begin{align}
    \log p(d_i| d_{-i}, \theta_{\rm GP}, \mathcal{M}) &= -\frac{1}{2} \left( \log 2 \pi \hat{\sigma}_i^2 + \frac{(r_i-\mu_i)^2}{\hat{\sigma}_i^2} \right), \\
    \hat{\sigma}^2_i &= \frac{1}{\left[ \bm{K}^{-1} \right]_{ii}}, \\
    \mu_i &= r_i - \frac{\left[ \bm{K}^{-1}r\right]_i}{\left[ \bm{K}^{-1} \right]_{ii}},
\end{align}
where $r_i=d_i-m_i$ is the residual of the model fit to the $i^{\rm th}$ data point and $\bm{K}^{-1}$ is the inverse of the covariance matrix. 

The PSIS approximation remains applicable in this case. The conditional predictive density simply replaces the pointwise likelihood in the importance ratio, so that $r_i^s = 1/p \left( d_i | d_{-i}, \theta^s, \mathcal{M} \right)$, and Equations \ref{eq:weights}--\ref{eq:psis} then follow unchanged. Therefore, once this calculation has been completed, the remainder of the LOO-CV method can be applied as described in Section \ref{subsec:loocv}. The Pareto shape parameter $k$ retains its role as a diagnostic. This approach to cross-validation for non-factorizable likelihoods is described in detail by \citet{Burkner2021}.


\section{Case Studies} \label{sec:results}

\subsection{Diagnosing mass accuracy and precision} \label{subsec:hd119}

\begin{figure*}
    \centering
    \includegraphics[width=\linewidth,trim={0 0 3cm 0},clip]{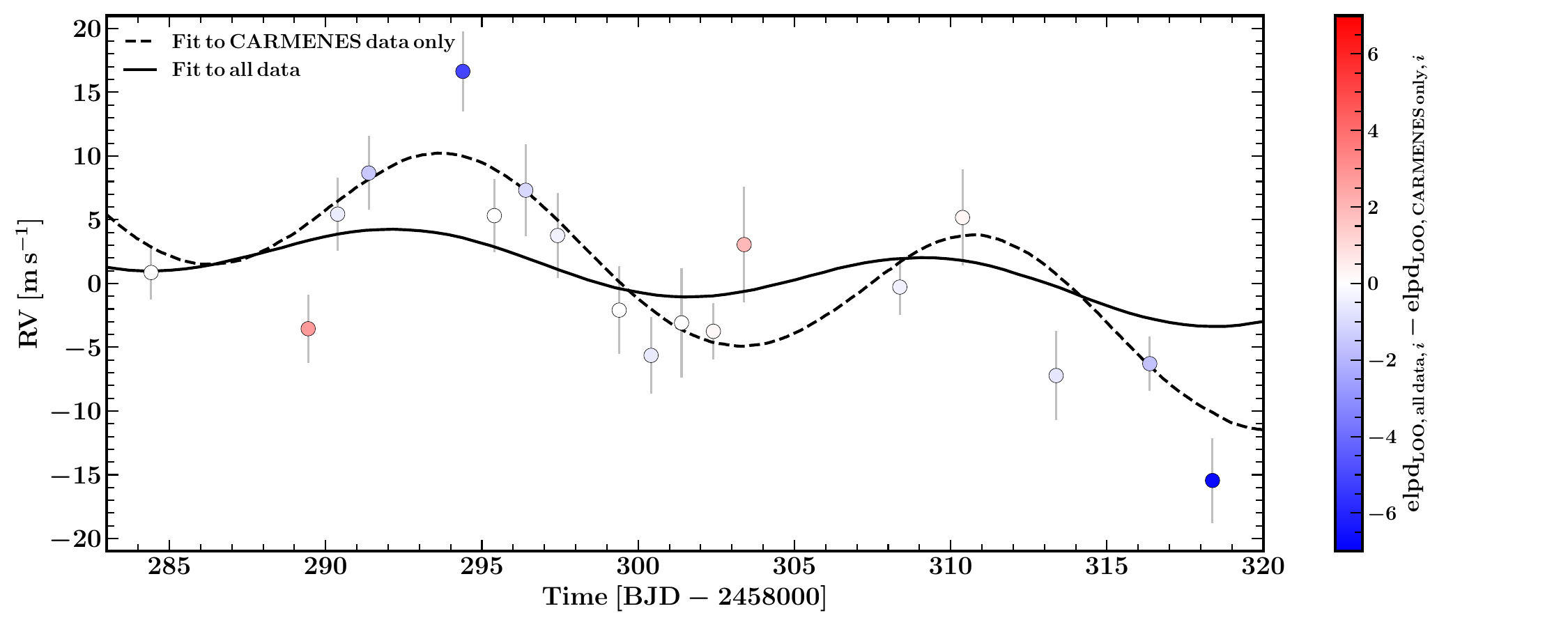}
    \caption{CARMENES RV observations of the HD~119130 system presented in \citet{Luque2019b}. The solid line shows the best-fit model to a data set combining the CARMENES observations with observations from HARPS, HARPS-N, and HIRES \citep[data from instruments other than CARMENES not shown, but can be seen in][]{AkanaMurphy2025}. The dashed line shows the best-fit model to the CARMENES observations only. Points are colored by the value of $\Delta\mathrm{elpd}_{\mathrm{LOO},\mathcal{M}_1,\mathcal{M}_2, i}$, where $\mathcal{M}_1$ is the model fit to the data from all instruments and $\mathcal{M}_2$ is the model fit to the CARMENES data only. Red points are better predicted by $\mathcal{M}_1$ when withheld, whereas blue points are better predicted by $\mathcal{M}_2$.}
    \label{fig:hd119_rv}
\end{figure*}

Our first case study explores how LOO-CV can be used to assess the reliability of a mass measurement obtained from a limited RV data set, using the example of HD~119130~b (also known as K2-292~b). RV measurements of HD~119130~b were acquired in 2018 using the CARMENES instrument \citep[Calar Alto high-Resolution search for M dwarfs with Exoearths with Near-infrared and optical Echelle Spectrographs,][]{Quirrenbach2014,Quirrenbach2018}. Analysis of 18 RVs (shown in Figure \ref{fig:hd119_rv}) resulted in an inferred planet mass $M_p = 24.5 \pm 4.4\,M_{\oplus}$. Combined with a planet radius $R_p=2.63^{+0.11}_{-0.10}\,R_{\oplus}$ measured from K2 light curves, this indicated an unusually high density $\rho_p = 7.4^{+1.6}_{-1.5}\,$g$\,$cm$^{-3}$, one of the highest values measured for a sub-Neptune up to that point, causing the planet to be labeled as ``ultradense'' \citep{Luque2019b}.

Follow-up observations of the target were subsequently taken using HARPS \citep[High Accuracy Radial velocity Planet Searcher,][]{Mayor2003}, HARPS-N \citep{Cosentino2012} and Keck/HIRES \citep[HIgh Resolution Echelle Spectrometer,][]{Vogt1994}. In total, 57 new RVs of HD~119130 were acquired, resulting in a combined 75 RVs for the system, analyzed in \citet{AkanaMurphy2025}. The analysis of the complete set of observations found a much smaller $M_p = 8.8 \pm 3.2\,M_{\oplus}$ than the analysis of the CARMENES data alone in \citet{Luque2019b}. The resulting planetary density $\rho_p=2.7 \pm 1.0\,$g$\,$cm$^{-3}$ is much more in line with other planets with $R_p<3\,R_{\oplus}$, indicating that HD~119130~b is not truly an ultradense sub-Neptune, but rather a planet with a typical density given its radius.

In this case study, we investigate the CARMENES RV data that were used to infer an anomalously high mass for HD~119130~b. We begin by revisiting the findings of both \citet{Luque2019b} and \citet{AkanaMurphy2025} and recovering the strong discrepancy in the measured value of $K$ from the analysis of the full set of 75 RVs compared to the CARMENES data alone. There are some small differences between the priors adopted in the two previous studies, and so for this study we define a common set of priors that are used for both the CARMENES-only fit and the joint fit. All priors and resulting posterior estimates are summarized in Table~\ref{Table:HD119130}. Our prior on the planetary orbital period $\mathcal{N}(16.9841, 0.001)$ d is similar to the best-fit solution reported by \citet{AkanaMurphy2025}. We choose an eccentricity prior $\beta(1.52, 29.0)$ that favors low eccentricities while still allowing the exploration of higher values \citep{2019AJ....157...61V}. Our prior for the time of periastron passage is $\mathcal{U}(2460445.0 \: \rm{d}, 2460462.0 \: \rm{d})$ or $\mathcal{U}(2458309.9 \: \rm{d}, 2458326.9 \: \rm{d})$, depending on whether the fit was performed on the full data set or CARMENES data alone. 

We measure $K=6.0^{+1.3}_{-1.4}\,$m$\,$s$^{-1}$ when fitting the CARMENES data alone, and $K=2.4^{+0.8}_{-0.8}$~m$\,$s$^{-1}$ when fitting the entire data set, finding very close agreement with the results of \citet{Luque2019b} and \citet{AkanaMurphy2025}, and recovering the result that a higher planetary mass is found when fitting only the CARMENES data compared to the full data set. 

We apply LOO-CV for each CARMENES data point in both cases. This involves calculating $\mathrm{elpd}_{\mathrm{LOO}, \mathcal{M}, i}$ for each of the 18 CARMENES RVs for both cases, requiring 36 calculations in total. Due to the relatively small number of calculations and low computational cost of fitting these data, we do not use the PSIS approximation in this case study, instead calculating $\mathrm{elpd}_{\mathrm{LOO}, \mathcal{M}, i}$ directly by refitting the data with each point removed one-at-a-time (see Equation \ref{eq:exact_loo}). We subsequently calculate $\Delta\mathrm{elpd}_{\mathrm{LOO},\mathcal{M}_1,\mathcal{M}_2, i}$, where $\mathcal{M}_1$ is the model fit to the data from all instruments and $\mathcal{M}_2$ is the model fit to the CARMENES data only.

Figure \ref{fig:hd119_rv} shows the CARMENES RVs for HD~119130, alongside the best-fit models from fits to the entire multi-instrument data set as well as to the CARMENES data alone. For each RV, the value of $\Delta\mathrm{elpd}_{\mathrm{LOO},\mathcal{M}_1,\mathcal{M}_2, i}$ is shown. It is evident that two data points have a particularly large negative value, favoring the high-mass solution: the RVs at BJD$-$2458000$=$294.4~d and 318.4~d are the only points where $|\Delta\mathrm{elpd}_{\mathrm{LOO},\mathcal{M}_1,\mathcal{M}_2, i}|>5$ ($-5.1$ and $-6.7$, respectively, compared to a mean $|\Delta\mathrm{elpd}_{\mathrm{LOO},\mathcal{M}_1,\mathcal{M}_2, i}|$ of 1.3 across the 18 CARMENES RVs), suggesting that these points are strongly driving the preference for a higher inferred semi-amplitude when only the CARMENES data are considered.

\begin{figure}
    \centering
    \includegraphics[width=\linewidth]{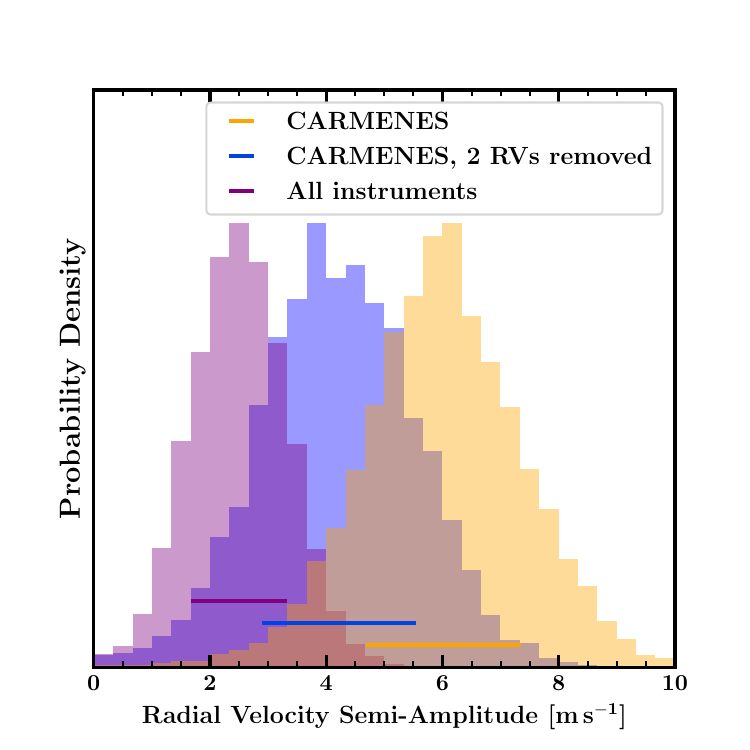}
    \caption{Posterior probability distributions for the RV semi-amplitude $K$ of the HD~119130 system. The orange posterior shows the result from modeling the CARMENES data alone. The blue posterior shows the result from modeling the CARMENES data with the two points with the largest negative values of $\Delta\mathrm{elpd}_{\mathrm{LOO},\mathcal{M}_1,\mathcal{M}_2, i}$ removed. The purple posterior shows the results from modeling the entire multi-instrument data set. Removing the two CARMENES RVs identified above significantly reduces the tension between the CARMENES-only and multi-instrument results.}
    \label{fig:k_post}
\end{figure}

To further investigate the impact of these two data points on the inferred properties of the system, we refit the original CARMENES data set, but leaving these two points out of the analysis. This results in a measured $K=4.2^{+1.4}_{-1.3}\,$m$\,$s$^{-1}$, a value much more closely in line with the fit to the entire data set (see Figure \ref{fig:k_post}). This indicates that these two CARMENES RVs are responsible for much of the tension between the inferred masses from each study.

In order to conclude that a given inference is robust, the results should not depend very strongly on just one or two data points. In this case, the need to reassess the CARMENES data set became clear when follow-up observations found a different planetary mass. However, in the case where further observations are unavailable, we recommend testing the sensitivity of inferred semi-amplitudes to the removal of individual data points (LOO-CV) or small subsets of the data (e.g., via $k$-fold cross validation). For HD~119130~b, the two points driving the larger mass inference were those with the largest positive and negative RV measurements. We suggest refitting any given RV data set with the largest positive and negative points removed to ensure that the inferred planetary properties do not change significantly. Furthermore, the RV at BJD$-$2458000$=$318.4~d was taken with the highest airmass out of all of the CARMENES observations. As an additional robustness test, we recommend refitting the data with those points corresponding to the worst observing conditions removed. We further discuss the benefits of additional tests in Section \ref{subsec:limitations}.

\subsection{Selecting between high-eccentricity and multi-planet models} \label{subsec:gj4276}

Here we ask which of two competing explanations for an observed RV signal is better supported by the data, and which future observations would best distinguish between them. For this case study, we examine the GJ~4276 system, for which RV measurements were obtained between 2016 and 2018 using CARMENES \citep{Nagel2019}. These observations led to the discovery of a Neptune-mass exoplanet with a very high eccentricity ($e=0.37$). However, \citet{Nagel2019} noted that the RV observations could be explained either by a single planet on an eccentric orbit or by a pair of planets on circular orbits, near a period ratio of 2:1 --- a well-known degeneracy arising from the Fourier expansion of the Kepler equation in powers of the eccentricity \citep[e.g.,][]{Moulton1914, Anglada-Escude2010}. \citet{Nagel2019} conducted a likelihood ratio test \citep{Wilks1938,Protassov2002} and found a slight preference for the eccentric single-planet solution. Despite this preference, the reported difference in likelihood was not sufficient to reject either solution with high confidence.

In this work, we conduct a LOO-CV analysis of observations of this system with two main goals. The first is to calculate the total $\mathrm{elpd}_{\mathrm{LOO}}$ for the single-planet and two-planet models, and examine the difference between these metrics to compare their predictive performance. This provides a complementary assessment of model preference, which can be used in conjunction with the likelihood ratio test from \citet{Nagel2019}. The second goal is to determine which data points are responsible for driving any preference between the two models. This potentially allows us to determine the orbital phases at which RV measurements have the most power in discriminating between models. This would enable targeted RV observations at those specific phases, allowing for more confident conclusions about the nature of the planetary system to be reached in an efficient manner.

We begin by reanalyzing the data presented in \citet{Nagel2019}. We consider a single-planet Keplerian fit and a two-planet circular model. The eccentricity and the argument of periastron are included as free parameters in the single-planet model and fixed to 0 and 90$^{\circ}$ respectively in the two-planet model. All priors and posterior estimates are listed in Table~\ref{Table:GJ4276}. Overall, we find very close agreement between our results and those reported in \citet{Nagel2019}, particularly given that a different RV model and sampling algorithm were used in that work.

We subsequently perform LOO-CV on both the single-planet and two-planet models. In contrast to the previous case study, we use the PSIS approximation so that $\mathrm{elpd}_{\mathrm{LOO}, \mathcal{M}, i}$ can be calculated without the need to perform individual refits withholding each data point one-at-a-time. This is particularly advantageous for the GJ~4276 data set, which consists of 100 RV observations, meaning a traditional LOO-CV analysis without using the PSIS approximation would require an additional 200 refits to the data to compare the two models at the per-data-point level.

\begin{figure}
    \centering
    \includegraphics[width=\linewidth]{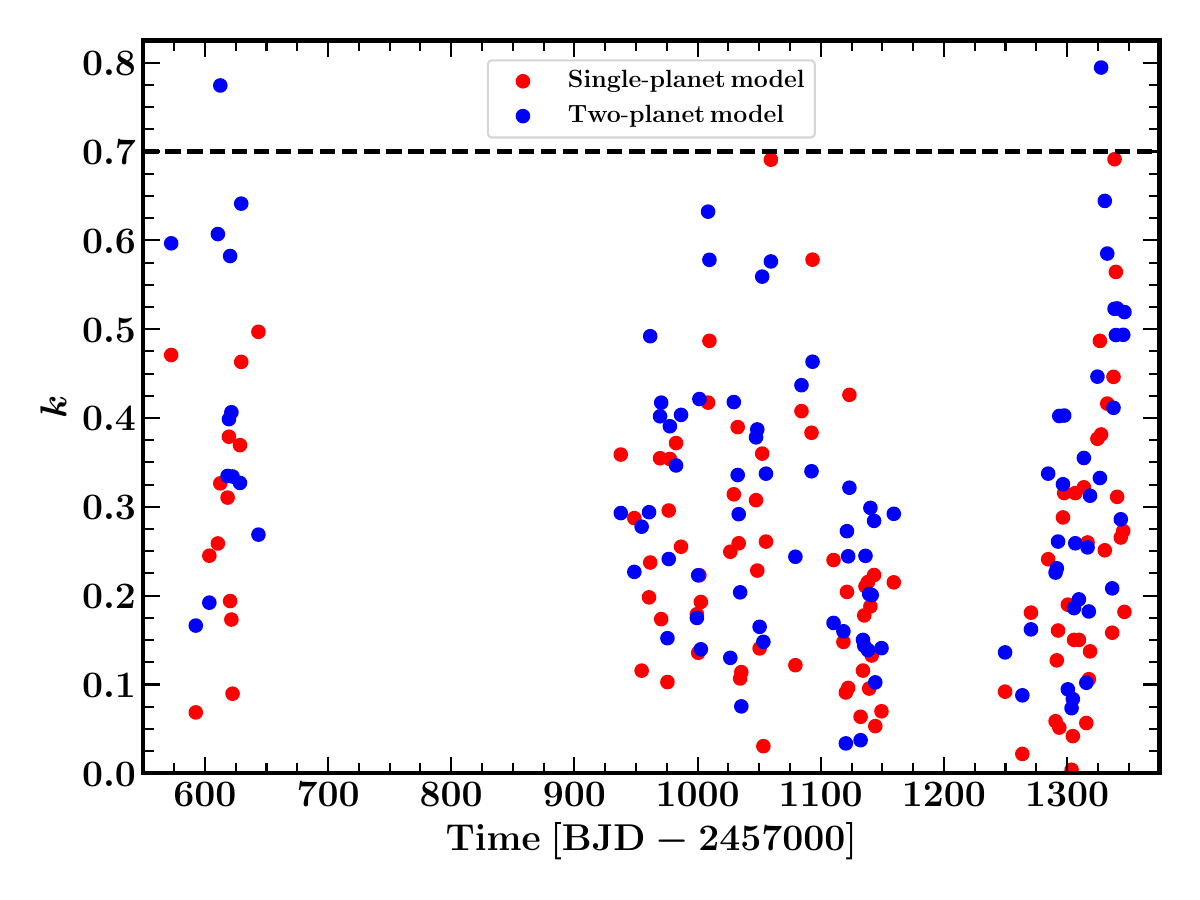}
    \caption{Estimated shape parameter $k$ of the Pareto distributions fit to each data point after conducting LOO-CV on the GJ~4276 observations using the PSIS approximation.}
    \label{fig:pareto_k}
\end{figure}

Since we use the PSIS approximation, we inspect the values of the shape parameter $k$ for each data point and for each model, shown in Figure \ref{fig:pareto_k}. We find $k<0.7$ for the vast majority of data points, meaning that in most cases, the $\mathrm{elpd}_{\mathrm{LOO}, \mathcal{M}, i}$ values from the PSIS approximation can be trusted. For the single-planet model, $k$ does not exceed 0.7 for any data points. For the two-planet model, $k>0.7$ for two data points (at BJD$-$2457000$=$612.5~d and 1327.6~d). We therefore conduct two additional fits of the data set using the two-planet model, leaving out each of those data points in turn. For those points, we calculate $\mathrm{elpd}_{\mathrm{LOO}, \mathcal{M}, i}$ following the same method as in Section \ref{subsec:hd119}. We ultimately carry out four fits to the data in total, rather than 202, which would be required if the PSIS approximation was not used.

\begin{figure*}
    \centering
    \includegraphics[width=\linewidth,trim={0 0 3cm 0},clip]{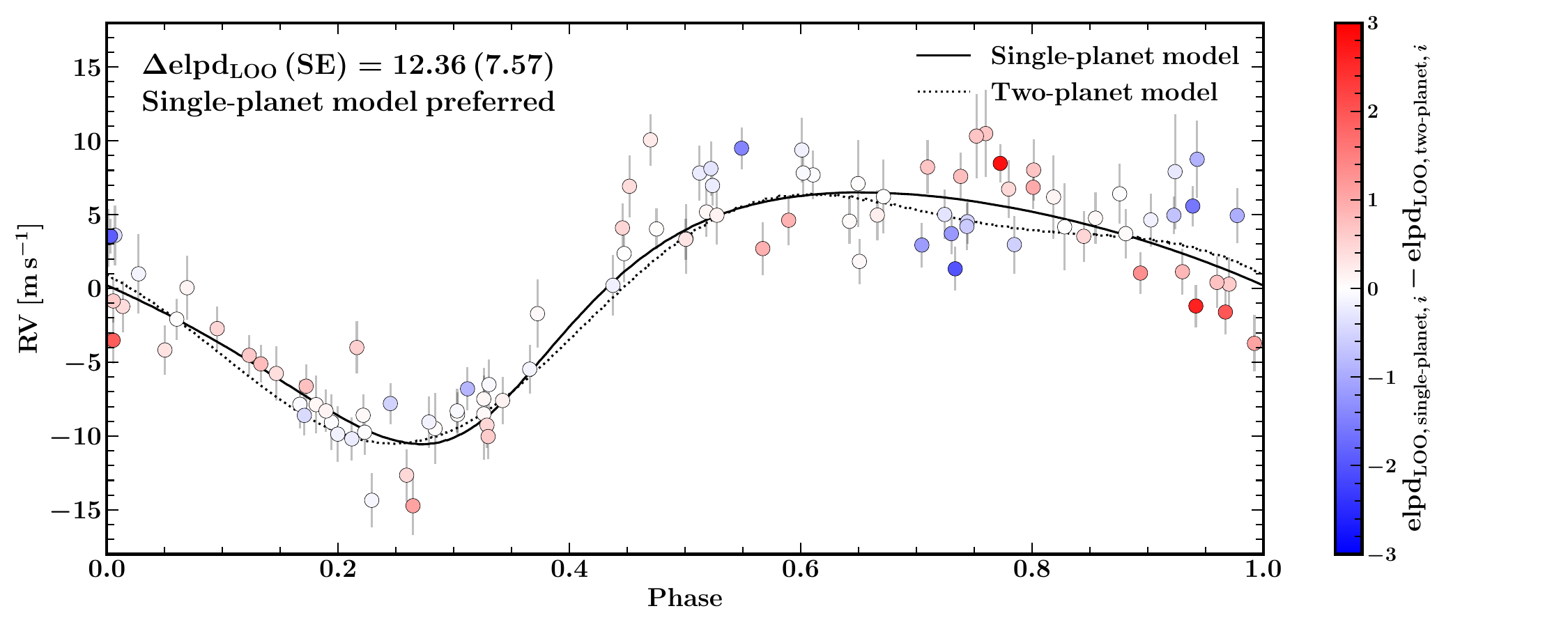}
    \caption{Phase-folded CARMENES RV observations of the GJ~4276 system. The solid line shows the best-fit model to the data assuming that the system consists of a single (eccentric) planet. The dotted line shows the best-fit model for a two-planet system, phase-folded to match the period of the single-planet model. Points are colored by the value of $\Delta\mathrm{elpd}_{\mathrm{LOO},\mathcal{M}_1,\mathcal{M}_2, i}$, where $\mathcal{M}_1$ is the single-planet model and $\mathcal{M}_2$ is the two-planet model. Red points are better predicted by $\mathcal{M}_1$ when withheld, whereas blue points are better predicted by $\mathcal{M}_2$. The total $\Delta\mathrm{elpd}_{\mathrm{LOO},\mathcal{M}_1,\mathcal{M}_2}$ and its standard error (SE) are also shown, revealing a preference for the single-planet model.}
    \label{fig:gj4276_rv}
\end{figure*}

Phase-folded CARMENES RVs for the GJ~4276 system are shown in Figure \ref{fig:gj4276_rv}, alongside best-fit realizations from the single-planet and two-planet models. Data points are colored by the value of $\Delta\mathrm{elpd}_{\mathrm{LOO},\mathcal{M}_1,\mathcal{M}_2, i}$, where $\mathcal{M}_1$ is the single-planet model and $\mathcal{M}_2$ is the two-planet model. For these two models, the total $\Delta\mathrm{elpd}_{\mathrm{LOO},\mathcal{M}_1,\mathcal{M}_2}=12.36$, with a standard error of 7.57, resulting in $\Delta\mathrm{elpd}_{\mathrm{LOO},\mathcal{M}_1,\mathcal{M}_2}/\mathrm{SE}=1.63$. The out-of-sample performance of the single-planet model is better than that of the two-planet model, indicating that the single-planet model is preferred by this metric. 

We note that the value of $\Delta\mathrm{elpd}_{\mathrm{LOO},\mathcal{M}_1,\mathcal{M}_2}/\mathrm{SE}$ has not been mapped to a qualitative degree of confidence for a model preference, as has been attempted for other model comparison metrics such as Bayes factors \citep[e.g.,][]{Kass1995,Trotta2008}. While we caution against the use of such mappings for this quantity, we encourage its use as a complementary metric alongside other model comparison tests. In this case, the preference for the single-planet model acts in support of the likelihood ratio test conducted by \citet{Nagel2019}, adding further credence to the hypothesis that the RV signal shown by the GJ~4276 system is the result of a single, eccentric planet.

\begin{figure}
    \centering
    \includegraphics[width=\linewidth]{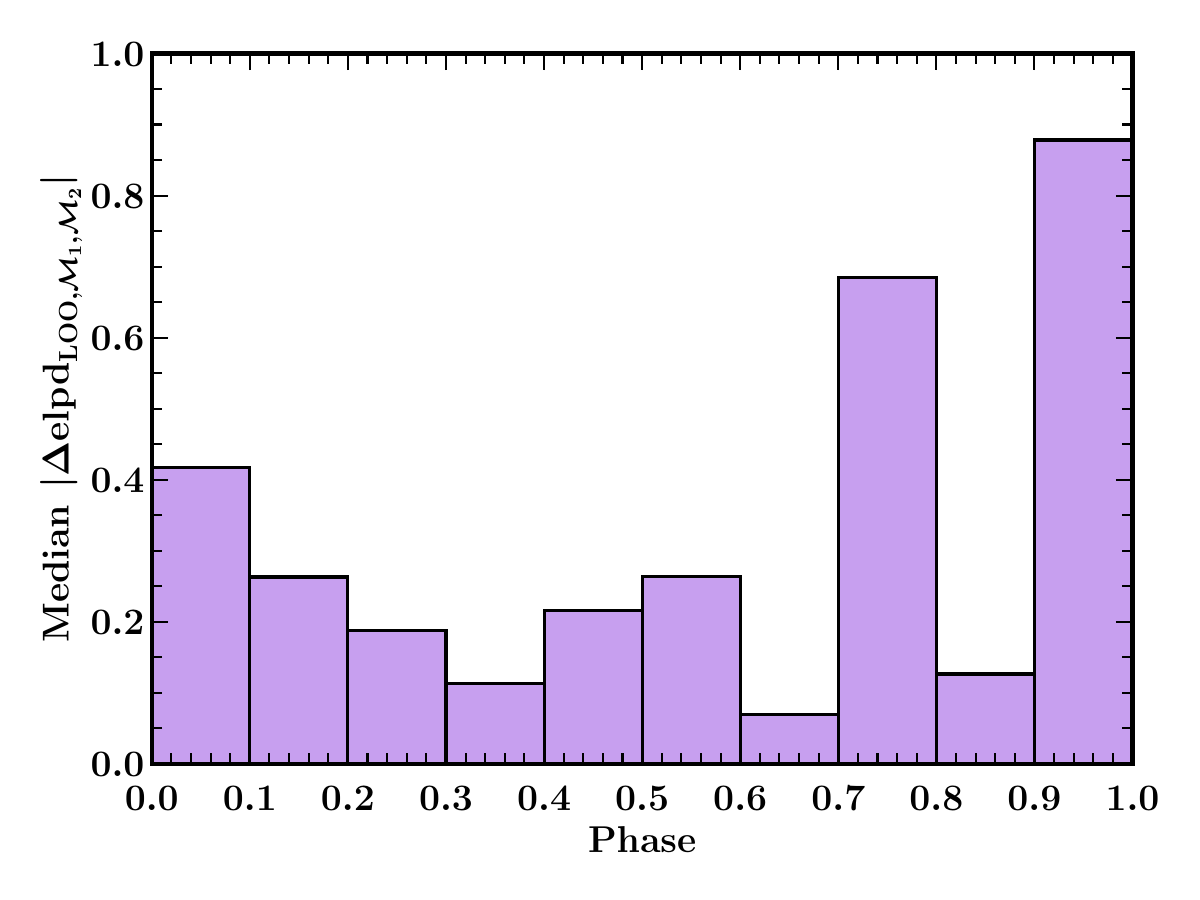}
    \caption{Binned median values of $|\Delta\mathrm{elpd}_{\mathrm{LOO},\mathcal{M}_1,\mathcal{M}_2, i}|$ between the single-planet and two-planet models of the GJ~4276 system, using phase bin widths of 0.1.}
    \label{fig:binned_dloo}
\end{figure}

In addition to providing a complementary metric of overall model preference, an advantage of model comparison using LOO-CV is that it reveals which data points, and when applied to RV data, which orbital phases are most responsible for driving any model preference. Figure \ref{fig:gj4276_rv} suggests that $|\Delta\mathrm{elpd}_{\mathrm{LOO},\mathcal{M}_1,\mathcal{M}_2, i}|$ is largest at orbital phases $\gtrsim0.7$. To quantify this apparent trend, we bin the values of $|\Delta\mathrm{elpd}_{\mathrm{LOO},\mathcal{M}_1,\mathcal{M}_2, i}|$ with a bin width of 0.1 in orbital phase, and calculate the median value within each bin. The resulting values are shown in Figure \ref{fig:binned_dloo}. We find that the highest median value of $|\Delta\mathrm{elpd}_{\mathrm{LOO},\mathcal{M}_1,\mathcal{M}_2, i}|$ is found at phases [0.9,1.0] (median of 0.88), followed by [0.7,0.8] (median of 0.68). For all other bins, the median is less than 0.5. This suggests that observations taken at phases [0.9,1.0] and [0.7,0.8] provide the most information when discriminating between models. This information could be used to guide follow-up observations, enabling efficient use of future observing time by taking measurements at the phases demonstrated to yield the highest predictive power. We note that the optimization of RV observing strategies, and of the choice of observing phases in particular, has been studied extensively \citep[e.g.,][]{Loredo2004,Ford2008,Baluev2008,BalsalobreRuza2023}. The quantity $|\Delta\mathrm{elpd}_{\mathrm{LOO},\mathcal{M}_1,\mathcal{M}_2, i}|$ tells us which measurements most strongly discriminate between two specified models, and is obtained directly from the data already in hand at no computational cost beyond the LOO-CV calculation itself. We therefore present it as a convenient and interpretable complement to existing scheduling methods.

\subsection{Assessing a marginal signal in the presence of correlated noise} \label{subsec:gj357}

Our final case study demonstrates LOO-CV for a model that includes correlated noise, using the GJ~357 system. GJ~357 is a nearby (9.4~pc) M2.5\,V dwarf hosting three known planets \citep{Luque2019}: a transiting, Earth-sized planet GJ~357~b ($P=3.93$~d), and two non-transiting planets GJ~357~c ($P=9.12$~d) and GJ~357~d ($P=55.7$~d). A periodogram analysis of the RVs by \citet{Luque2019} revealed a fourth periodicity at $\sim87.3$~d. Among purely Keplerian models, a four-planet model attained the highest Bayesian evidence of any model they tested. However, the inferred orbit of the fourth planet was substantially more eccentric than those of the other planets, and the authors noted that a GP with an exponential kernel accounted for the same $87.3$~d periodicity. They therefore adopted a model consisting of three Keplerian orbits along with a GP. Here we ask whether LOO-CV supports the inclusion of a fourth planet in this system.

We use the data set of 141 RVs presented by \citet{Luque2019}, comprising 32 measurements from Keck/HIRES \citep{Vogt1994}, 30 from VLT/UVES \citep{Zechmeister2009}, 53 from HARPS \citep{Mayor2003}, 16 from Magellan/PFS \citep{Crane2010}, and 10 from CARMENES. The PFS measurements taken before and after the February 2018 slit change are treated as separate instruments (PFSpre and PFSpost, comprising 7 and 9 measurements, respectively), since they require different velocity offsets.

We fit models comprising one, two, three and four Keplerian signals, in each case with the periods restricted to the four periodicities identified by \citet{Luque2019} and with circular orbits. Following \citet{Luque2019}, we include a GP with an exponential kernel,

\begin{equation}
    \hat{k}_{i,j} = \sigma_{\rm GP}^2 \exp \left( - |t_i - t_j| / t_{\rm GP} \right),
    \label{eq:expkernel}
\end{equation}

\noindent shared across all instruments, where $\sigma_{\rm GP}$ is an amplitude and $t_{\rm GP}$ is a characteristic timescale, together with one velocity offset and one jitter term for each instrument.

Priors and posterior estimates are given in Table \ref{Table:GJ357}. The inferred GP amplitude decreases from $\sigma_{\rm GP}=2.5$ (one-planet model) to $1.1\,$m$\,$s$^{-1}$ (four-planet model) as Keplerian signals are added to the model, illustrating the reassignment of variance from the noise model to the planetary signals.

\begin{figure}
    \centering
    \includegraphics[width=\linewidth]{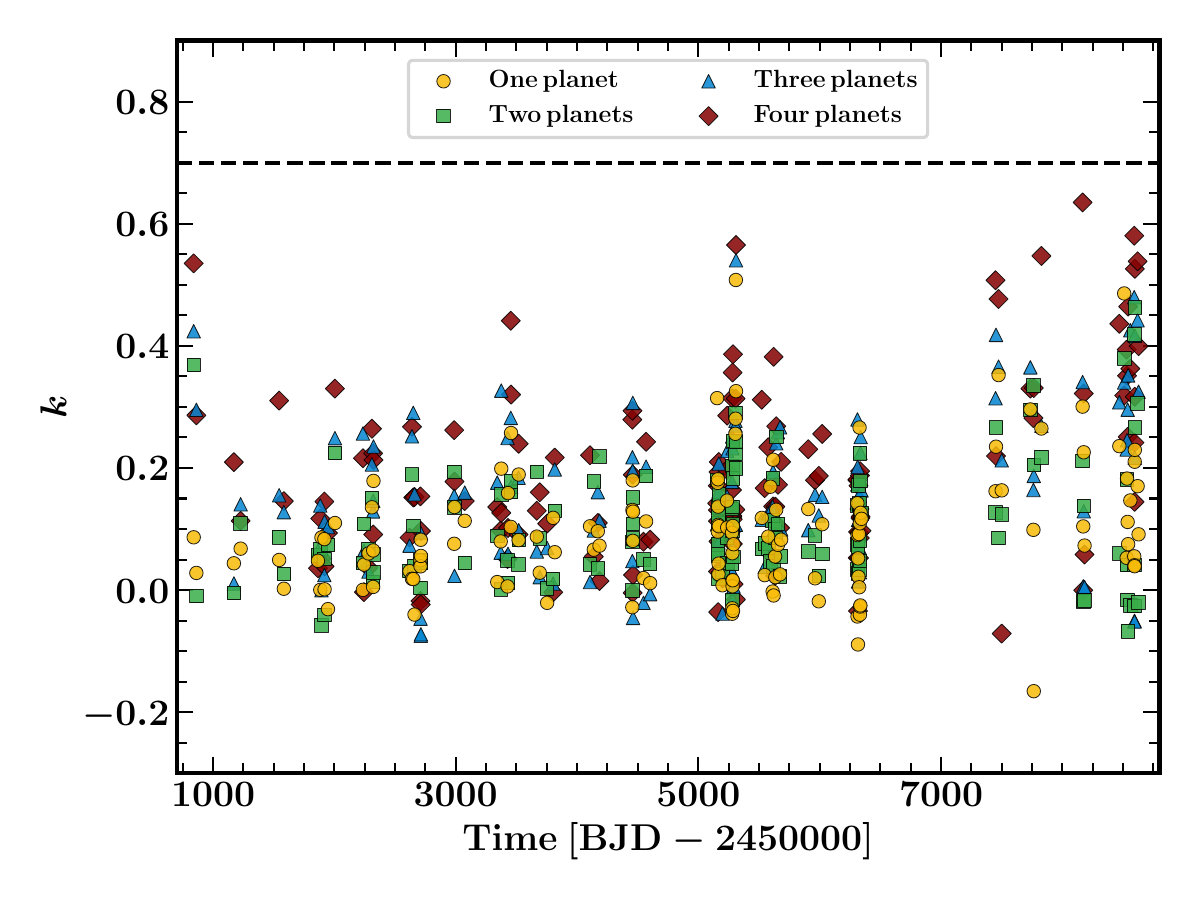}
    \caption{Estimated shape parameter $k$ of the Pareto distributions fit to each data point after conducting LOO-CV on the GJ~357 observations, for each of the four models. No data point exceeds the threshold of $k=0.7$ in any model, so no exact refits are required.}
    \label{fig:gj357_pareto_k}
\end{figure}

We compute $\mathrm{elpd}_{\mathrm{LOO}, \mathcal{M}, i}$ for each of the 141 RVs and each of the four models using the conditional predictive density from Section \ref{subsec:corrnoise}, together with the PSIS approximation. The Pareto shape parameter estimates for each data set and each model are shown in Figure \ref{fig:gj357_pareto_k}. It does not exceed 0.7 for any data point in any of the four models, reaching a maximum of 0.63, so the PSIS approximation is reliable throughout and no exact refits are required. The complete analysis therefore requires four fits to the data, rather than the 568 that would be needed for direct LOO-CV.

Results of the model comparison are summarized in Table \ref{Table:GJ357_comparison}. The two-planet model is clearly preferred over the one-planet model, with $\Delta\mathrm{elpd}_{\mathrm{LOO}}/\mathrm{SE}=4.29$, and the three-planet model is preferred over the two-planet model, with $\Delta\mathrm{elpd}_{\mathrm{LOO}}/\mathrm{SE}=2.04$. In contrast, adding a fourth Keplerian yields $\Delta\mathrm{elpd}_{\mathrm{LOO}}=0.63$ against a standard error of 3.17, i.e., $\Delta\mathrm{elpd}_{\mathrm{LOO}}/\mathrm{SE}=0.20$, indicating that the four-planet model provides no improvement in out-of-sample predictive performance over the three-planet model. Furthermore, stacking of the predictive distributions \citep{Yao2018,Nixon2024} assigns weights of 44\% and 56\% to the three- and four-planet models, suggesting that neither model is favored. The Bayesian evidence, in contrast, moderately favors the four-planet model ($\Delta \ln \mathcal{Z} = 3.4$).

\begin{table*}
    \centering
    \begin{tabular}{c|c|c|c|c|c}
    Model & $\mathrm{elpd}_{\mathrm{LOO}}$ (SE) & $\ln \mathcal{Z}$ & $\Delta\ln \mathcal{Z}$ & $\Delta\mathrm{elpd}_{\mathrm{LOO}}/\mathrm{SE}$ & $w_{\rm stack}$ \\
    \hline
    1 planet  & $-370.2 \: (9.1)$ & $-391.3$ & - & - & 0.00 \\
    2 planets & $-346.3 \: (8.4)$ & $-381.7$ & 9.64 & 4.29 & 0.00 \\
    3 planets & $-339.4 \: (8.3)$ & $-373.4$ & 8.36 & 2.04 & 0.44 \\
    4 planets & $-338.8 \: (8.8)$ & $-369.9$ & 3.41 & 0.20 & 0.56 \\
    \hline
    \end{tabular}
    \caption{Model comparison for the GJ~357 system. $\Delta\mathrm{elpd}_{\mathrm{LOO}}/\mathrm{SE}$ compares each model with the next simplest model in the sequence, using the standard error from Equation \ref{eq:se_2}. $w_{\rm stack}$ is the weight assigned to each model by stacking of the predictive distributions \citep{Yao2018}.}
    \label{Table:GJ357_comparison}
\end{table*}

\begin{figure*}
    \centering
    \includegraphics[width=\linewidth]{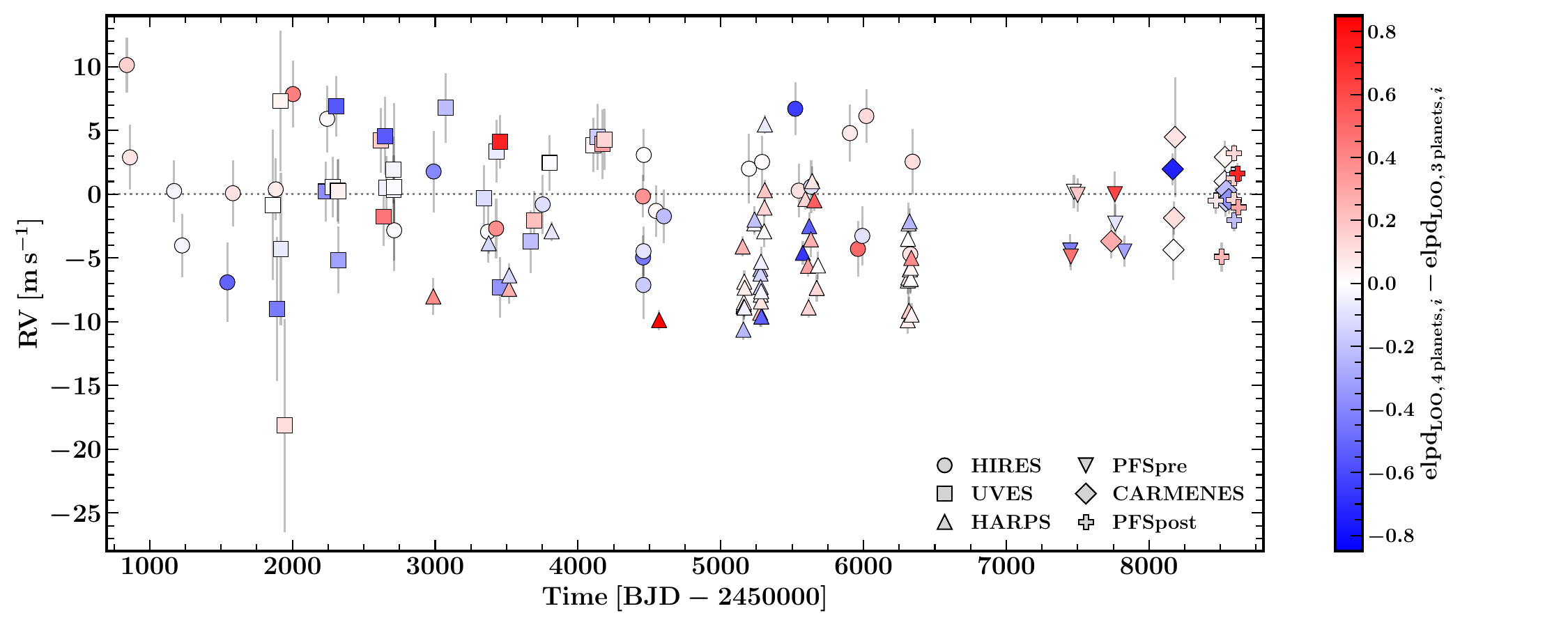}
    \caption{RV observations of the GJ~357 system from \citet{Luque2019}. Marker shape indicates the instrument. Points are colored by the value of $\Delta\mathrm{elpd}_{\mathrm{LOO},\mathcal{M}_1,\mathcal{M}_2, i}$, where $\mathcal{M}_1$ is the four-planet model and $\mathcal{M}_2$ is the three-planet model. Red points are better predicted by $\mathcal{M}_1$, whereas blue points are better predicted by $\mathcal{M}_2$. Note that the color scale spans a much smaller range than in Figure \ref{fig:hd119_rv}.}
    \label{fig:gj357_rv}
\end{figure*}

Figure \ref{fig:gj357_rv} shows the RV time series, with each point colored by $\Delta\mathrm{elpd}_{\mathrm{LOO},\mathcal{M}_1,\mathcal{M}_2, i}$ where $\mathcal{M}_1$ is the four-planet model and $\mathcal{M}_2$ the three-planet model. No individual measurement strongly prefers either model: the values span $-0.73$ to $+0.83$, with a mean $|\Delta\mathrm{elpd}_{\mathrm{LOO},\mathcal{M}_1,\mathcal{M}_2, i}|$ of 0.19, and 72 of the 141 points favor the four-planet model, consistent with the $\sim$50\% expected if the two models predict the data equally well. This contrasts with Figure \ref{fig:hd119_rv}, where individual points reach $|\Delta\mathrm{elpd}_{\mathrm{LOO},\mathcal{M}_1,\mathcal{M}_2, i}|>5$.

\begin{figure}
    \centering
    \includegraphics[width=\linewidth]{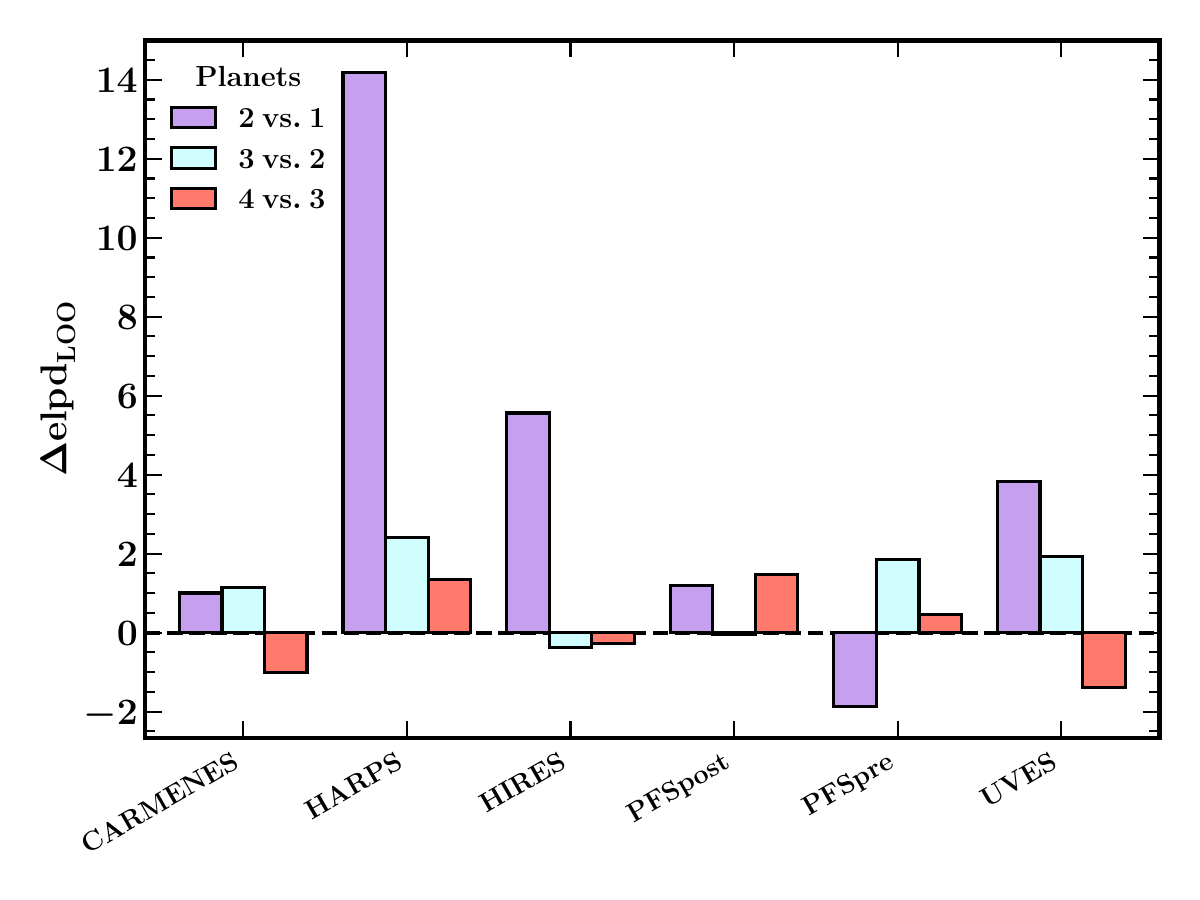}
    \caption{Contribution of each instrument to $\Delta\mathrm{elpd}_{\mathrm{LOO},\mathcal{M}_1,\mathcal{M}_2}$ between different models of the GJ~357 system.}
    \label{fig:gj357_per_inst}
\end{figure}

Because these observations come from six data sets (five instruments, with PFSpost and PFSpre treated separately), it is also informative to decompose $\Delta\mathrm{elpd}_{\mathrm{LOO},\mathcal{M}_1,\mathcal{M}_2}$ by data set, as shown in Figure \ref{fig:gj357_per_inst}. The improvement from one to two planets is supported by five of the six data sets, and that from two to three planets by four of six (and only very weakly disfavored by the other two). The fourth Keplerian, however, is supported by some data sets and disfavored by others: for example, $\Delta\mathrm{elpd}_{\mathrm{LOO}}=+1.4$ for HARPS and $+1.5$ for PFSpost, and $-1.4$ for UVES and $-1.0$ for CARMENES. This means that the fourth planet signal lacks multi-instrument support.

\begin{figure*}
    \centering
    \includegraphics[width=\linewidth]{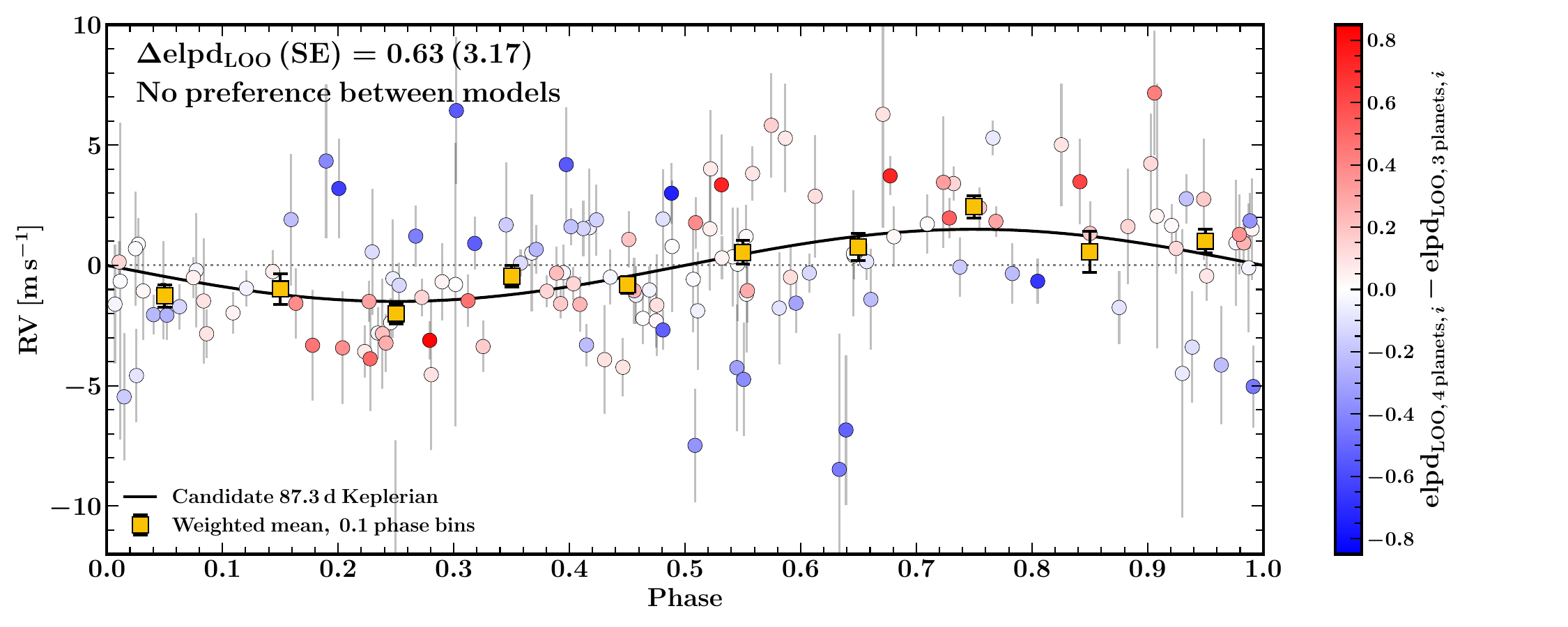}
    \caption{RV observations of GJ~357, phase-folded on the candidate 87.3~d signal, after subtracting the Keplerian signals of the three known planets and the instrumental velocity offsets at the posterior medians of the four-planet fit. The solid line shows the candidate Keplerian from the same fit. Points are colored as in Figure \ref{fig:gj357_rv}. Gold squares show weighted means in bins of 0.1 in phase. Note that the uncertainties on these binned means assume independent individual measurements, and are therefore underestimated: the fitted GP has a correlation timescale of 4.5~d, so measurements falling in the same phase bin are correlated.}
    \label{fig:gj357_folded}
\end{figure*}

Figure \ref{fig:gj357_folded} shows the same measurements phase-folded on the period of the potential fourth planet (87.3~d), after subtracting the three known planets and the instrumental offsets. The binned weighted means (weighted by measurement uncertainty) do broadly follow the candidate Keplerian, and a na\"{i}ve $\chi^2$ comparison of those binned values would strongly prefer the Keplerian over a flat line ($\chi^2 = 15.3$ for the Keplerian versus 74.5 for a flat line, for 10 phase bins.) However, the uncertainties on the binned means are propagated assuming that the measurements are independent, which is not the case here. With a correlation timescale of 4.5~d (the posterior median for the four-planet model), measurements falling within the same phase bin are correlated, and the uncertainties on the binned means are correspondingly underestimated. The LOO-CV analysis, which accounts for this correlation through the full covariance matrix, provides no predictive support for a fourth planetary signal in the GJ~357 system.

\section{Discussion and Summary} \label{sec:discussion}

\subsection{Practical recommendations for application to RV analysis}

Here, we summarize best practices for applying LOO-CV as a model interrogation tool for RV analysis.

In cases where a single RV model has been fit to a single data set, LOO-CV can be used to identify outlier points that are strongly driving the inference. Here, two distinct quantities are of interest, which diagnose different things. The estimated shape parameter $k$ of the Pareto distribution diagnoses the reliability of the PSIS computation: if $k \geq 0.7$ for any data points, the approximation cannot be trusted for those points and an exact refit with each such point removed is required in order to evaluate $\mathrm{elpd}_{\mathrm{LOO}, \mathcal{M}, i}$. The pointwise values of $\mathrm{elpd}_{\mathrm{LOO}, \mathcal{M}, i}$, by contrast, diagnose the influence of individual measurements on the inference. Both can motivate a refit, but for different reasons: in the first case to obtain a reliable LOO-CV estimate, and in the second to establish whether the inferred parameters depend strongly on a small number of measurements. These methods should be combined with checking the airmass and seeing conditions of any outlier points as an additional cross-check of their reliability.

Inspection of Figure \ref{fig:hd119_rv} by eye indicates that several of the CARMENES measurements lie some distance from the best-fit models, and such an inspection is a valuable first check on any RV data set. The advantage of LOO-CV is that it quantifies and ranks the influence of each measurement, rather than simply identifying those that appear discrepant. It is not apparent by eye, for example, whether the measurement at BJD$-$2458000$=$289.4~d is more or less influential than the one at 313.4~d, whereas LOO-CV shows that the former has a larger effect on the inferred semi-amplitude. This distinction becomes more important for the larger data sets considered in Sections \ref{subsec:gj4276} and \ref{subsec:gj357}, where the most influential measurements are not visually conspicuous: the measurement that contributes most to the model preference in the GJ~4276 data set lies at an orbital phase of 0.77 and would not obviously have been identified by eye.

In cases where competing models have been fit to data, LOO-CV can be informative about both the data and the models. As shown in Section \ref{subsec:hd119}, LOO-CV can be used to interrogate changes in inferred results when a model initially fit to a small data set is fit to additional data. In this case, both the estimated shape parameter $k$ (if PSIS is used) and $\Delta\mathrm{elpd}_{\mathrm{LOO},\mathcal{M}_1,\mathcal{M}_2}$ are valuable metrics to determine the reliability of the original data set, where anomalous values of either parameter may justify a refit.

Additionally, LOO-CV can be used for overall model comparison, as demonstrated in Sections \ref{subsec:gj4276} and \ref{subsec:gj357}. In this case, the metric of interest is $\Delta\mathrm{elpd}_{\mathrm{LOO},\mathcal{M}_1,\mathcal{M}_2}/\mathrm{SE}$. This should be used in conjunction with other model comparison tools, such as the Bayesian evidence and likelihood ratio. We do not advocate for any specific ``threshold'' at which a particular model should be accepted over another, but we note that disagreement between metrics serves as an indicator that model preferences are not reliable for the given analysis. For the GJ~357 system, Bayesian evidence moderately favors a fourth Keplerian signal while LOO-CV shows no improvement in predictive performance, strengthening the three-planet architecture reported by \citet{Luque2019}.

It is crucial to bear in mind that any model performance metric has its own strengths and drawbacks. For example, the Bayesian evidence can be highly sensitive to the chosen prior, which may result in biases if used to compare models \citep{Gelman2013}. Since LOO-CV is performed by integrating over the posterior distribution (see Equation \ref{eq:prob}), it is less susceptible to this pitfall. LOO-CV is not entirely free of prior dependence, since the posterior itself depends on the prior, but it is considerably less sensitive to prior choice than the evidence. Where the aim is to weight models rather than to select one of them, model weights such as those obtained by stacking of predictive distributions (for which LOO-CV is required) provide a further alternative \citep{Yao2018}.

Finally, LOO-CV can be used to identify optimal timing for follow-up observations in order to effectively discriminate between competing models. For this goal, the values of $|\Delta\mathrm{elpd}_{\mathrm{LOO},\mathcal{M}_1,\mathcal{M}_2, i}|$ should be used. Specific times or orbital phases where $|\Delta\mathrm{elpd}_{\mathrm{LOO},\mathcal{M}_1,\mathcal{M}_2, i}|$ is large indicate that observations at those phases are more powerful for distinguishing between models. Therefore, timing follow-up observations to occur at those times is likely to yield the most additional information in order to determine the preferred model.

Where a marginal signal is inferred from a compilation of observations from several instruments, we further recommend decomposing $\Delta\mathrm{elpd}_{\mathrm{LOO},\mathcal{M}_1,\mathcal{M}_2}$ by instrument, as demonstrated in Section \ref{subsec:gj357}. Any real planetary signal should improve the prediction of measurements from independent instruments, whereas a preference that is confined to a single data set, or that changes sign between data sets, may serve as a warning that the inference is not robust.

We note that the PSIS approximation, while useful for improving the efficiency of LOO-CV analysis, is more likely to fail for smaller data sets (where failure is manifested as $k \geq 0.7$ for a significant fraction of the data). In cases where the number of data points is small, and model fitting is inexpensive, we recommend a na\"{i}ve approach to LOO-CV where each data point is left out in turn.

To aid with community adoption of LOO-CV methods into RV analysis, we present an online tutorial which will enable the reader to reproduce the results shown in Sections \ref{subsec:gj4276} and \ref{subsec:gj357} for the GJ~4276 and GJ~357 systems. The tutorial is hosted on GitHub\footnote{\url{https://github.com/mcnixon/rv\_loo\_cv\_tutorial}} and preserved on Zenodo\footnote{\href{https://doi.org}{doi:10.5281/zenodo.22779651}}.

\subsection{Limitations and future directions} \label{subsec:limitations}

Leave-one-out cross-validation is not the only cross-validation scheme available, and for time series it is not always the most appropriate one. When the noise is correlated on timescales comparable to the spacing between observations, a withheld measurement can be predicted in part by interpolating between its immediate temporal neighbors, which limits the sensitivity of LOO-CV to structure varying on longer timescales. Leave-future-out and block cross-validation schemes, in which contiguous subsets of the time series are withheld, avoid this and are better suited to assessing model components that vary slowly \citep{Burkner2020}. The formalism of Section \ref{subsec:corrnoise} extends naturally to these schemes, since the joint conditional predictive density of a withheld block is likewise available analytically.

Additionally, the exponential kernel adopted in Section \ref{subsec:gj357} describes stochastic variability but not rotationally modulated activity. Applying LOO-CV with quasi-periodic kernels, and using it to compare activity models rather than only planetary architectures, is a natural next step. In cases where GP regression is required, we recommend the use of additional metrics such as the MGIC$_{\rm rv}$ \citep{Barragan2026} to further assess model performance.

We note that LOO-CV should not be considered the only tool to assess model performance, but rather one of several methods that can be employed to determine the reliability of a given result. In the case of HD~119130~b described in Section \ref{subsec:hd119}, LOO-CV was able to identify the data points driving an unusually high mass, after additional observations were acquired that were in tension with the CARMENES data. If LOO-CV is instead applied only to the CARMENES data (i.e., computing $\mathrm{elpd}_{\mathrm{LOO}, \mathcal{M}, i}$ for the high-mass solution alone), we find that the point at BJD$-$2458000$=$294.4~d is the second most poorly predicted of the 18, with $\mathrm{elpd}_{\mathrm{LOO}, \mathcal{M}, i}=-5.9$ against a median of $-2.5$. The measurement at 318.4~d, however, is not distinctive, with a value of $-3.2$. As the largest negative RV in the sample, this point is responsible for driving the fit to a higher mass, and therefore it is well-predicted by the high-mass solution. This illustrates a limitation of LOO-CV: when applied to a single data set, it can identify which measurements are not well-predicted by the model, but those measurements are not necessarily the same as those which determine a given parameter. We therefore recommend the combination of LOO-CV with other tests to determine reliability, as described in Section \ref{subsec:hd119}.

\subsection{Summary and concluding remarks}

In this work, we presented a LOO-CV analysis framework for RV observations. We implemented the PSIS approximation to enable efficient LOO-CV calculations without the need to refit each data point in turn, which is particularly valuable for relatively large data sets ($\gtrsim$100 measurements). We applied LOO-CV to identify outliers in the CARMENES RV observations of HD~119130~b, showing that the anomalously large mass inferred from those observations was largely driven by just two data points. Upon re-inspection of the data, one of these points was also found to have a high airmass. We also demonstrated how LOO-CV can be applied to model selection with the case of the GJ~4276 system, revealing a preference for an eccentric single-planet model over a two-planet model. Furthermore, we showed that in this case, LOO-CV can be used to identify specific windows of time during which additional observations would be most powerful for distinguishing between competing models. Thirdly, we extended LOO-CV to RV models containing correlated noise, and applied it to the GJ~357 system, whose observations are modeled with a GP. In this case LOO-CV provides no predictive support for a fourth planetary signal that is moderately favored by the Bayesian evidence, strengthening the three-planet architecture reported by \citet{Luque2019} and illustrating the value of assessing marginal signals with more than one model comparison metric. We outlined practical guidance for implementing LOO-CV in RV analysis. In each case, LOO-CV allowed us to ask not only which model best describes the data, but which individual measurements drive the results, and whether the resulting inference is robust.

We believe that LOO-CV is a useful tool that can form part of the arsenal of statistical analysis methods used to maximize the information gained from existing and future RV data sets. By refining statistical methodologies to improve both precision and accuracy, this framework can help to optimize observing strategies, improve the reliability of inferred planetary properties, and ensure that mass constraints remain robust for next-generation exoplanet surveys which will push the field to new limits.

\begin{acknowledgments}

The authors thank Peter McGill, with support from the Lawrence Livermore National Laboratory Space Science Institute, for contributing to the initial ideas that led to this paper and to the interpretation of some of the results. We also thank Finnegan Keller for helpful feedback on the manuscript and online tutorial, and the anonymous referee for their careful review of our work. M.N. and L.W. thank the Heising-Simons Foundation for their funding through the 51 Pegasi b Postdoctoral Fellowship. Support for this work was provided by NASA to R.L. through the NASA Hubble Fellowship grant awarded by the Space Telescope Science Institute, which is operated by the Association of Universities for Research in Astronomy, Inc., for NASA, under contract NAS5-26555. R.L. is funded by the European Union (ERC, THIRSTEE, 101164189) and acknowledges financial support from the Severo Ochoa grant CEX2021-001131-S funded by MCIN/AEI/10.13039/501100011033. Views and opinions expressed are however those of the author(s) only and do not necessarily reflect those of the European Union or the European Research Council. Neither the European Union nor the granting authority can be held responsible for them. M.N. and L.W. acknowledge support from NASA XRP Grant 80NSSC24K0160. This research has made use of the NASA Astrophysics Data System and the NASA Exoplanet Archive.

\end{acknowledgments}

\software{\textsc{NumPy} \citep{Harris2020},   
          \textsc{Matplotlib} \citep{Hunter2007},
          \textsc{ArviZ} \citep{ArviZ2019},
          \textsc{juliet} \citep{Juliet},
          \textsc{RadVel} \citep{2018PASP..130d4504F},
          \textsc{dynesty} \citep{Speagle2020}
          }

\appendix

\renewcommand{\thetable}{A.\arabic{table}}
\setcounter{table}{0}

\section{Planetary and Instrumental Parameters}

\subsection{HD~119130}\label{app:hd}

\begin{table}
    \centering
    \begin{tabular}{c|c|c}
    Parameter (unit) & Estimate & Prior \\
    \hline 
    \multicolumn{3}{c}{Fit to CARMENES Data Only}\\
    \hline
    \multicolumn{3}{c}{Fitted Parameters}\\
    \hline
    Orbital Period $P$ (d) & $16.984 \pm 0.001$ & $\mathcal{N}(16.9841, 0.001)$ \\
    RV Semi-Amplitude $K$ (m\,s$^{-1}$) & $6.03^{+1.29}_{-1.36}$ & $\mathcal{U}(0, 20)$ \\
    Eccentricity $e$ & $0.044^{+0.053}_{-0.028}$ & $\beta(1.52, 29.0)$ \\
    Argument of Periastron $\omega$ ($^{\circ}$) & $-15.59^{+109.66}_{-92.24}$& $\mathcal{U}(-180, 180)$ \\
    Time of Periastron Passage $t_{0}$ (d) & $2458315.3^{+0.6}_{-0.7}$ & $\mathcal{U}(2458309.9, 2458326.9)$ \\
    Velocity Offset $\gamma_{\rm CARMENES}$ (m\,s$^{-1}$) & $0.94^{+13.04}_{-13.97}$ & $\mathcal{U}(-20, 20)$ \\
    Instrumental Jitter $\sigma_{\rm CARMENES}$ (m\,s$^{-1}$) & $1.84^{+1.64}_{-1.49}$ & $\mathcal{LU}(0.1, 20)$\\
    RV Slope (m\,s$^{-1}$\,d$^{-1}$) & $-0.39 \pm 0.09$ & $\mathcal{U}(-2, 2)$ \\ 
    RV Intercept (m\,s$^{-1}$) & $-61.85^{+18.73}_{-19.51}$ & $\mathcal{U}(-200, 200)$ \\
    \hline
    \multicolumn{3}{c}{Derived Parameters}\\
    \hline
    Planetary Mass ($M_{\oplus}$) & 24.15 $\pm$ 5.65 & - \\
    \hline
    \multicolumn{3}{c}{}\\
    \multicolumn{3}{c}{Fit to All Instruments}\\
    \hline
    \multicolumn{3}{c}{Fitted Parameters}\\
    \hline
    Orbital Period $P$ (d) & $16.984 \pm 0.001$ & $\mathcal{N}(16.9841, 0.001)$ \\
    RV Semi-Amplitude $K$ (m\,s$^{-1}$) & $2.44^{+0.84}_{-0.79}$ & $\mathcal{U}(0, 20)$ \\
    Eccentricity $e$ & $0.041^{+0.046}_{-0.026}$ & $\beta(1.52, 29.0)$ \\
    Argument of Periastron $\omega$ ($^{\circ}$) & $22.86^{+112.73}_{-151.06}$ & $\mathcal{U}(-180, 180)$ \\
    Time of Periastron Passage $t_{0}$ (d) & $2460454.35^{+0.86}_{-0.74}$ & $\mathcal{U}(2460445.0,2460462.0)$ \\
    Velocity Offset $\gamma_{\rm CARMENES}$ (m\,s$^{-1}$) & $-10.52^{+10.96}_{-6.88}$ & $\mathcal{U}(-20, 20)$ \\
    Velocity Offset $\gamma_{\rm HARPS}$ (m\,s$^{-1}$) & $8.03^{+6.39}_{-10.53}$ & $\mathcal{U}(-20, 20)$ \\
    Velocity Offset $\gamma_{\rm HARPS-N}$ (m\,s$^{-1}$) & $10.64^{+6.71}_{-11.39}$ & $\mathcal{U}(-20, 20)$ \\
    Velocity Offset $\gamma_{\rm HIRES}$ (m\,s$^{-1}$) & $8.87^{+6.22}_{-10.57}$ & $\mathcal{U}(-20, 20)$ \\
    Instrumental Jitter $\sigma_{\rm CARMENES}$ (m\,s$^{-1}$) & $5.34^{+1.48}_{-1.16}$ & $\mathcal{LU}(0.1, 20)$\\
    Instrumental Jitter $\sigma_{\rm HARPS}$ (m\,s$^{-1}$) & $4.30^{+0.86}_{-0.72}$ & $\mathcal{LU}(0.1, 20)$\\
    Instrumental Jitter $\sigma_{\rm HARPS-N}$ (m\,s$^{-1}$) & $3.71^{+1.22}_{-0.99}$ & $\mathcal{LU}(0.1, 20)$\\
    Instrumental Jitter $\sigma_{\rm HIRES}$ (m\,s$^{-1}$) & $1.55^{+0.92}_{-0.88}$ & $\mathcal{LU}(0.1, 20)$\\
    RV Slope (m\,s$^{-1}$\,d$^{-1}$) & $-0.009^{+0.006}_{-0.005}$ & $\mathcal{U}(-2, 2)$ \\ 
    RV Intercept (m\,s$^{-1}$) & $9.62^{+6.60}_{-10.22}$ & $\mathcal{U}(-200, 200)$ \\
    \hline
    \multicolumn{3}{c}{Derived Parameters}\\
    \hline
    Planetary Mass ($M_{\oplus}$) & 9.78 $\pm$ 3.34 & - \\
    \hline
    \end{tabular}
    \caption{Prior distributions and posterior estimates for fits to RV observations of the HD~119130 system (consisting of a single planet, HD~119130~b). Results are shown for the fit to the CARMENES data alone \citep{Luque2019b}, as well as to the multi-instrument data set \citep{AkanaMurphy2025}. Posterior estimates reflect the median value and $\pm1\sigma$ uncertainties for each parameter. $\mathcal{N}(\cdot,\cdot)$ indicates a Gaussian distribution, $\mathcal{U}(\cdot,\cdot)$ indicates a uniform distribution, $\mathcal{LU}(\cdot,\cdot)$ indicates a log-uniform distribution, and $\beta(\cdot,\cdot)$ indicates a $\beta$ distribution.}
    \label{Table:HD119130}
\end{table}

\subsection{GJ 4276}\label{app:gj4276}

\begin{table}
    \centering
    \begin{tabular}{c|c|c}
    Parameter (unit) & Estimate & Prior \\
    \hline 
    \multicolumn{3}{c}{Single-Planet (Keplerian Orbit) Model}\\
    \hline
    \multicolumn{3}{c}{Fitted Parameters}\\
    \hline
    Orbital Period $P_{b}$ (d) & $13.3512^{+0.0049}_{-0.0053}$ & $\mathcal{U}(13.1, 13.5)$ \\
    RV Semi-Amplitude $K_{b}$ (m\,s$^{-1}$) & $8.49^{+0.36}_{-0.37}$ & $\mathcal{U}(0, 100)$ \\
    Time of Periastron Passage $t_{0,b}$ (d) & $2458347.62^{+0.24}_{-0.25}$ & $\mathcal{U}(2458339.84, 2458353.24)$ \\
    Eccentricity $e_{b}$ & $0.301^{+0.039}_{-0.042}$ & $\beta(1.52, 29.0)$ \\
    Argument of Periastron $\omega_{b}$ ($^{\circ}$) & $-144.56^{+7.65}_{-7.58}$ & $\mathcal{U}(-180, 180)$ \\
    Velocity Offset $\gamma_{\rm CARMENES}$ (m\,s$^{-1}$) & $0.40 \pm 0.25$ & $\mathcal{U}(-50, 50)$ \\
    Instrumental Jitter $\sigma_{\rm CARMENES}$ (m\,s$^{-1}$) & $1.73^{+0.28}_{-0.25}$ & $\mathcal{LU}(0.1, 20)$ \\
    \hline
    \multicolumn{3}{c}{Derived Parameters}\\
    \hline
    $M_{b}\sin i$ ($M_{\oplus}$) & 16.31 $\pm$ 0.88 & - \\
    \hline
    \multicolumn{3}{c}{}\\
    \multicolumn{3}{c}{Two-Planet (Circular Orbit) Model}\\
    \hline
    \multicolumn{3}{c}{Fitted Parameters}\\
    \hline
    Orbital Period $P_{b}$ (d) & $13.3504^{+0.0059}_{-0.0061}$ & $\mathcal{U}(13.1, 13.5)$ \\
    RV Semi-Amplitude $K_{b}$ (m\,s$^{-1}$) & $7.61 \pm 0.36$ & $\mathcal{U}(0, 100)$ \\
    Time of Periastron Passage $t_{0,b}$ (d) & $2458346.42^{+0.16}_{-0.15}$ & $\mathcal{U}(2458339.84, 2458353.24)$ \\
    Eccentricity $e_{b}$ & - & 0 \\
    Argument of Periastron $\omega_{b}$ ($^{\circ}$) & - & $90$ \\
    Orbital Period $P_{c}$ (d) & $6.6744 \pm 0.0062$ & $\mathcal{U}(6.3, 6.9)$ \\
    RV Semi-Amplitude $K_{c}$ (m\,s$^{-1}$) & $2.66^{+0.35}_{-0.33}$ & $\mathcal{U}(0, 100)$ \\
    Time of Periastron Passage $t_{0,c}$ (d) & $2458348.59 \pm 0.31$ & $\mathcal{U}(2458343.19, 2458349.89)$ \\
    Eccentricity $e_{c}$ & - & 0 \\
    Argument of Periastron $\omega_{c}$ ($^{\circ}$) & - & $90$ \\
    Velocity Offset $\gamma_{\rm CARMENES}$ (m\,s$^{-1}$) & $0.33 \pm 0.26$ & $\mathcal{U}(-50, 50)$ \\
    Instrumental Jitter $\sigma_{\rm CARMENES}$ (m\,s$^{-1}$) & $1.85^{+0.27}_{-0.24}$ & $\mathcal{LU}(0.1, 20)$ \\
    \hline
    \multicolumn{3}{c}{Derived Parameters}\\
    \hline
     $M_{b}\sin i$ ($M_{\oplus}$) & 15.35 $\pm$ 0.91 & - \\
     $M_{c}\sin i$ ($M_{\oplus}$) & 4.25 $\pm$ 0.58 & - \\
     \hline
    \end{tabular}
    \caption{Prior distributions and posterior estimates for fits to CARMENES RV observations of the GJ~4276 system \citep{Nagel2019}. Results are shown for a single-planet model on a Keplerian orbit and a two-planet model with assumed circular orbits. Posterior estimates reflect the median value and $\pm1\sigma$ uncertainties for each parameter. $\mathcal{U}(\cdot,\cdot)$ indicates a uniform distribution, $\mathcal{LU}(\cdot,\cdot)$ indicates a log-uniform distribution, and $\beta(\cdot,\cdot)$ indicates a $\beta$ distribution.}
    \label{Table:GJ4276}
\end{table}

\subsection{GJ 357}\label{app:gj357}

\begin{table}
    \centering
    \begin{tabular}{c|c|c|c}
    Parameter (unit) & 3 planets & 4 planets & Prior \\
    \hline
    \multicolumn{4}{c}{Fitted Parameters}\\
    \hline
    $P_{b}$ (d) & $3.930790^{+0.000001}_{-0.000001}$ & $3.930790^{+0.000001}_{-0.000001}$ & $\mathcal{N}(3.93079, 10^{-6})$ \\
    $K_{b}$ (m\,s$^{-1}$) & $1.47 \pm 0.28$ & $1.58 \pm 0.28$ & $\mathcal{U}(0, 10)$ \\
    $t_{0,b}$ (d) & $2458517.9910^{+0.0098}_{-0.0097}$ & $2458517.9905^{+0.0100}_{-0.0097}$ & $\mathcal{N}(2458517.99, 0.01)$ \\
    $P_{c}$ (d) & $9.1246^{+0.0012}_{-0.0011}$ & $9.1245^{+0.0014}_{-0.0013}$ & $\mathcal{N}(9.1, 0.01)$ \\
    $K_{c}$ (m\,s$^{-1}$) & $2.11^{+0.30}_{-0.31}$ & $1.87 \pm 0.30$ & $\mathcal{U}(0, 10)$ \\
    $t_{0,c}$ (d) & $2458314.26^{+0.45}_{-0.39}$ & $2458314.25^{+0.53}_{-0.44}$ & $\mathcal{U}(2458312, 2458318)$ \\
    $P_{d}$ (d) & $55.664^{+0.058}_{-0.060}$ & $55.696^{+0.055}_{-0.056}$ & $\mathcal{N}(55.7, 0.25)$ \\
    $K_{d}$ (m\,s$^{-1}$) & $2.08 \pm 0.38$ & $1.97 \pm 0.30$ & $\mathcal{U}(0, 10)$ \\
    $t_{0,d}$ (d) & $2458326.15^{+4.14}_{-4.00}$ & $2458328.04^{+3.77}_{-3.89}$ & $\mathcal{U}(2458310, 2458340)$ \\
    $P_{\rm cand}$ (d) & - & $87.310^{+0.148}_{-0.134}$ & $\mathcal{N}(87.3, 0.25)$ \\
    $K_{\rm cand}$ (m\,s$^{-1}$) & - & $1.50^{+0.38}_{-0.39}$ & $\mathcal{U}(0, 10)$ \\
    $t_{0,\rm cand}$ (d) & - & $2458298.49^{+5.69}_{-4.97}$ & $\mathcal{U}(2458280, 2458340)$ \\
    $\sigma_{\rm GP}$ (m\,s$^{-1}$) & $1.63 \pm 0.31$ & $1.13^{+0.36}_{-0.45}$ & $\mathcal{LU}(\sqrt{0.1}, \sqrt{10})$ \\
    $t_{\rm GP}$ (d) & $8.42^{+11.73}_{-4.46}$ & $4.46^{+96.76}_{-3.04}$ & $\mathcal{LU}(0.1, 10^{4})$ \\
    $\gamma_{\rm CARMENES}$ (m\,s$^{-1}$) & $-1.67 \pm 0.97$ & $-2.38^{+0.93}_{-0.87}$ & $\mathcal{U}(-10, 10)$ \\
    $\gamma_{\rm HARPS}$ (m\,s$^{-1}$) & $-5.17^{+0.43}_{-0.45}$ & $-5.02^{+0.34}_{-0.36}$ & $\mathcal{U}(-10, 10)$ \\
    $\gamma_{\rm HIRES}$ (m\,s$^{-1}$) & $0.98 \pm 0.67$ & $0.68^{+0.68}_{-0.67}$ & $\mathcal{U}(-10, 10)$ \\
    $\gamma_{\rm PFSpost}$ (m\,s$^{-1}$) & $-1.06 \pm 1.05$ & $-1.03^{+0.87}_{-0.81}$ & $\mathcal{U}(-10, 10)$ \\
    $\gamma_{\rm PFSpre}$ (m\,s$^{-1}$) & $-2.25^{+1.15}_{-1.22}$ & $-2.55^{+1.05}_{-1.06}$ & $\mathcal{U}(-10, 10)$ \\
    $\gamma_{\rm UVES}$ (m\,s$^{-1}$) & $1.09^{+0.77}_{-0.73}$ & $1.40 \pm 0.75$ & $\mathcal{U}(-10, 10)$ \\
    $\sigma_{\rm CARMENES}$ (m\,s$^{-1}$) & $0.87^{+0.99}_{-0.61}$ & $1.16^{+1.07}_{-0.79}$ & $\mathcal{U}(0, 10)$ \\
    $\sigma_{\rm HARPS}$ (m\,s$^{-1}$) & $0.75^{+0.44}_{-0.46}$ & $0.93^{+0.43}_{-0.57}$ & $\mathcal{U}(0, 10)$ \\
    $\sigma_{\rm HIRES}$ (m\,s$^{-1}$) & $2.10 \pm 0.88$ & $2.32^{+0.77}_{-0.76}$ & $\mathcal{U}(0, 10)$ \\
    $\sigma_{\rm PFSpost}$ (m\,s$^{-1}$) & $1.20^{+1.15}_{-0.78}$ & $1.02^{+0.91}_{-0.65}$ & $\mathcal{U}(0, 10)$ \\
    $\sigma_{\rm PFSpre}$ (m\,s$^{-1}$) & $1.59^{+1.44}_{-1.04}$ & $1.58^{+1.43}_{-1.04}$ & $\mathcal{U}(0, 10)$ \\
    $\sigma_{\rm UVES}$ (m\,s$^{-1}$) & $1.73^{+0.97}_{-0.96}$ & $2.20^{+0.89}_{-0.95}$ & $\mathcal{U}(0, 10)$ \\
    \hline
    \multicolumn{4}{c}{Derived Parameters}\\
    \hline
    $M_{b}$ ($M_{\oplus}$) & $1.77 \pm 0.34$ & $1.90 \pm 0.34$ & - \\
    $M_{c} \sin i$ ($M_{\oplus}$) & $3.36^{+0.50}_{-0.49}$ & $2.98 \pm 0.48$ & - \\
    $M_{d} \sin i$ ($M_{\oplus}$) & $6.06^{+1.13}_{-1.12}$ & $5.74^{+0.90}_{-0.89}$ & - \\
    $M_{\rm cand} \sin i$ ($M_{\oplus}$) & - & $5.09^{+1.28}_{-1.33}$ & - \\
    \hline
    \end{tabular}
    \caption{Prior distributions and posterior estimates for fits to the RV observations of the GJ~357 system presented by \citet{Luque2019}. Results are shown for the three-planet and four-planet models. One- and two-planet models use the same priors where relevant. All eccentricities are fixed to zero. $P_{\rm cand}$, $K_{\rm cand}$ and $t_{0,\rm cand}$ refer to the candidate 87.3~d signal, which we do not claim as planetary (see Section \ref{subsec:gj357}). Posterior estimates reflect the median value and $\pm1\sigma$ uncertainties for each parameter. Minimum masses are derived from the posterior samples of $P$ and $K$ assuming $M_{\star} = 0.342 \pm 0.011\,M_{\odot}$ \citep{Luque2019}. Since GJ~357~b transits, $\sin i > 0.9998$ so we report its minimum mass as a true mass. $\mathcal{N}(\cdot,\cdot)$ indicates a Gaussian distribution, $\mathcal{U}(\cdot,\cdot)$ indicates a uniform distribution, and $\mathcal{LU}(\cdot,\cdot)$ indicates a log-uniform distribution. Note that in practice, \texttt{juliet} fits for $a=\sigma_{\rm GP}^2$ and $c=1/t_{\rm GP}$ rather than for $\sigma_{\rm GP}$ and $t_{GP}$ directly. Priors and posterior estimates shown here have been converted to reflect measurements of $\sigma_{\rm GP}$ and $t_{\rm GP}$.}
    \label{Table:GJ357}
\end{table}

\bibliography{main}{}
\bibliographystyle{aasjournalv7}



\end{document}